\documentclass[prl,amsmath,twocolumn, showpacs, superscriptaddress,10pt]{revtex4-1}
\usepackage{amssymb}
\usepackage{amsmath}
\usepackage{hyperref}
\usepackage{graphicx}

\usepackage{color}
\definecolor{Blue}{rgb}{0.00, 0.00, 1.00}
\definecolor{Red}{rgb}{1.00, 0.00, 0.00}

\newcommand{\be}{\begin{equation}}
\newcommand{\ee}{\end{equation}}
\newcommand{\beq}{\begin{eqnarray}}
\newcommand{\eeq}{\end{eqnarray}}

\begin{document}

\title{Universal ground state properties of free fermions in a $d$-dimensional trap}

\author{David S. \surname{Dean}}
\affiliation{Univ. Bordeaux and CNRS, Laboratoire Ondes et Mati\`ere  d'Aquitaine
(LOMA), UMR 5798, F-33400 Talence, France}
\author{Pierre Le Doussal}
\affiliation{CNRS-Laboratoire de Physique Th\'eorique de l'Ecole Normale Sup\'erieure, 24 rue Lhomond, 75231 Paris Cedex, France}
\author{Satya N. \surname{Majumdar}}
\affiliation{Univ. Paris-Sud, CNRS, LPTMS, UMR 8626, Orsay F-91405, France}
\author{Gr\'egory \surname{Schehr}}
\affiliation{Univ. Paris-Sud, CNRS, LPTMS, UMR 8626, Orsay F-91405, France}

\begin{abstract} 
The ground state properties of $N$ spinless free fermions in a $d$-dimensional confining potential are studied. We find that any $n$-point correlation function has a simple determinantal structure that allows us to compute several properties exactly for large $N$. We show that the average density has a finite support with an edge, and near this edge the density exhibits a universal (valid for a wide class of potentials) scaling behavior for large $N$. The associated edge scaling function is computed exactly and generalizes to any $d$ the edge electron gas result of Kohn and Mattsson in $d=3$ [Phys. Rev. Lett. {\bf 81}, 3487 (1998)]. In addition, we calculate the kernel (that characterizes any $n$-point correlation function) for large $N$ and show that, when appropriately scaled, it depends only on dimension $d$, but has otherwise universal scaling forms, at the edges. The edge kernel, for higher $d$, generalizes the Airy kernel in one dimension, well known from random matrix theory. 
\end{abstract}


\maketitle

Over the past few years, experimental developments in the construction of optical traps and cooling protocols have allowed the study of systems of confined utracold atoms \cite{BDZ08,GPS08}. Non-interacting fermions or bosons at low temperatures are of particular interest as they exhibit purely quantum effects. For example, Bose-Einstein condensation \cite{BDZ08,GPS08} has been observed experimentally in several cold atom systems. There are also nontrivial quantum effects in non-interacting fermionic atoms, arising purely from the Pauli exclusion principle. A well studied example is a system of $N$ spinless fermions in a one-dimensional harmonic potential, $V(x) = \frac{1}{2} m \omega^2 x^2$ \cite{gleisberg,einstein,calabrese_prl,vicari_pra,vicari_pra2,vicari_pra3,eisler_prl,marino_prl,castillo,CDM14}.  
At zero temperature, $T=0$, the many-body ground state wavefunction $\Psi_0(x_1, \cdots, x_N)$ can be easily computed from the Slater determinant yielding \cite{einstein,marino_prl,CDM14}
\begin{equation}
|\Psi_0(x_1, \cdots, x_N)|^2 = \frac{1}{z_N} \prod_{i<j} (x_i - x_j)^2 e^{-\alpha^2\sum_{i=1}^N x_i^2}\label{rm1},
\end{equation}  
where $\alpha = \sqrt{m\omega/\hbar}$ has the dimension of inverse length and $z_N$ is a normalization constant. The squared wavefunction $|\Psi_0(x_1, \cdots, x_N)|^2$, characterizing the quantum fluctuations at $T=0$, can then be interpreted as the joint distribution of the eigenvalues of an $N \times N$ random matrix belonging to the Gaussian Unitary Ensemble (GUE) \cite{mehta,forrester}. 
\begin{figure}[ht]
\includegraphics[width=0.9\linewidth]{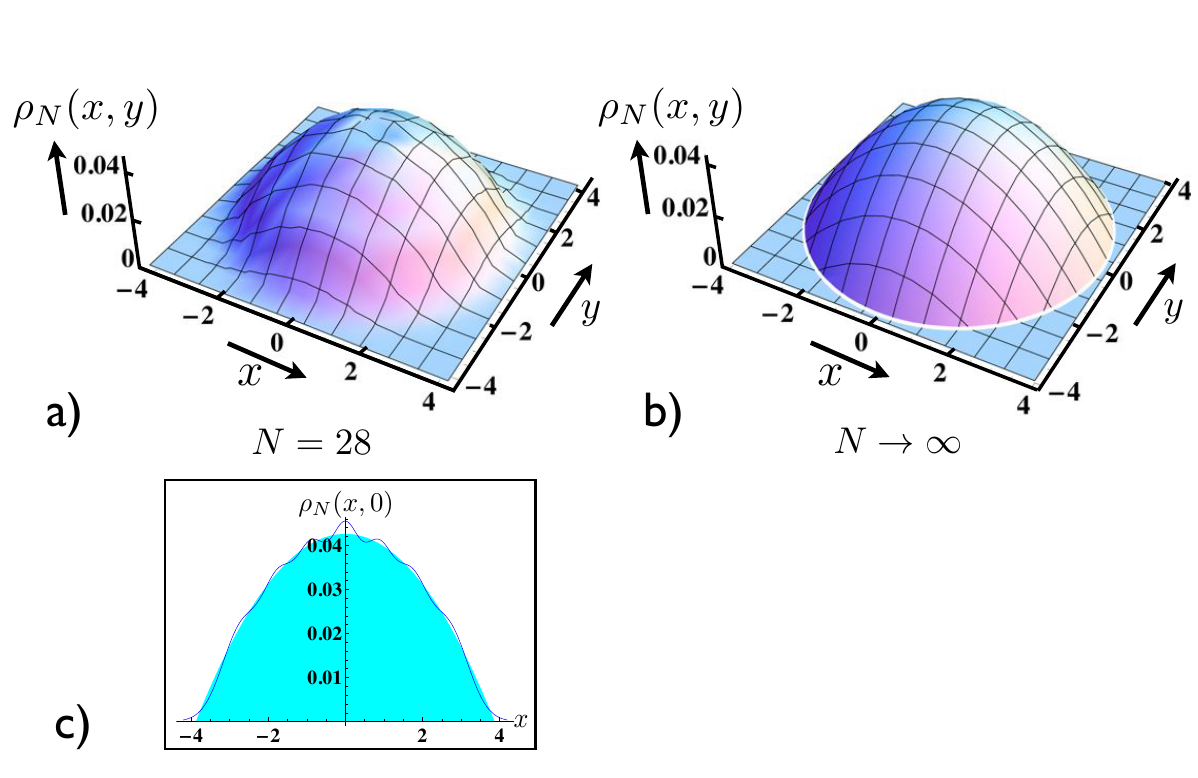}
\caption{(Color online) (a) Plot of $\rho_{N}(x,y)$ vs. $x$ and $y$ for $N=28$ fermions in a harmonic potential in $d=2$ obtained from the exact evaluation \cite{hermite} of Eqs. (\ref{eq:def_kernel}) and (\ref{eq:rel_density_kernel}). (b) Plot of $\rho_{N}(x,y)$ in the $(x,y)$ plane predicted from 
the asymptotic formula~Eq.~(\ref{eq:rho2}) with $V({\bf x}) = \frac{1}{2}m\omega^2 r^2$ and $\mu \approx \hbar \omega [\Gamma(d+1)\,N]^{1/d}$. (c) Comparison of (a) and (b) for $y=0$ (the region below the asymptotic result is shown via the  shaded region).}\label{bulkdensityshofig}
\end{figure}
Consequently, several zero temperature properties in this one-dimensional fermionic system have been computed analytically \cite{calabrese_prl,vicari_pra,vicari_pra2,vicari_pra3,eisler_prl,marino_prl,castillo,CDM14} using established results from random matrix theory (RMT). For example, the average density of fermions is given, for large $N$, by the celebrated Wigner semi-circular law \cite{mehta,forrester}:
\begin{eqnarray}\label{wigner}
\rho_N(x) \approx \frac{\alpha}{\sqrt{N}} f_W\left(\frac{\alpha \, x}{\sqrt{N}} \right) \;; \; f_W(z) = \frac{1}{\pi}\sqrt{2-z^2} 
\end{eqnarray}     
with sharp edges at $\pm \sqrt{2N}/\alpha$. These sharp edges get smeared for finite, but large, $N$ over a width $w_N \sim N^{-1/6}$ 
and the density near the edge (say, the right one) is described by a finite size scaling form \cite{BB91,For93}
\begin{eqnarray}\label{edge_density_1d}
\rho_N(x) \approx \frac{1}{N \, w_N} F_1\left[\frac{x - \sqrt{2N}/\alpha}{w_N} \right]
\end{eqnarray}  
with $w_N = N^{-1/6}/(\alpha \sqrt{2})$ and $F_1(z) = [Ai'(z)]^2 - z [Ai(z)]^2$, where $Ai(z)$ is the Airy function. Far to the left of the right edge, using $F_1(z)\sim \sqrt{|z|}/\pi$ as $z \to -\infty$, one can show that the scaling form (\ref{edge_density_1d}) smoothly matches with the semi-circular density in the bulk (\ref{wigner}). The edge scaling function $F_1(z)$ has been shown \cite{eisler_prl} to be universal, i.e., independent of the precise shape of the confining potential $V(x)$. In addition, all $n$-point correlation functions at zero temperature can be expressed as determinants constructed from a fundamental quantity called the kernel (see later for its precise definition). For large $N$, and away from the edge, the appropriately scaled kernel converges \cite{eisler_prl} to the universal sine-kernel form known from RMT \cite{mehta,forrester}. In contrast, near the edges, it approaches \cite{eisler_prl} the universal Airy kernel, also well known from RMT \cite{For93,TW}. Recently, for one-dimensional traps, several of the zero temperature results for the density as well as the kernel have been extended to finite temperature \cite{finiteT}.

In many experimental setups, the optical traps are actually in higher dimensions $d>1$. For $d>1$, there is no obvious relation between free fermions in a trap and RMT. Consequently, calculating analytically the zero temperature properties of spinless free fermions in a $d$-dimensional confining potential is a challenging problem. The bulk properties can be  estimated rather accurately using the local density approximation (LDA, also known as the Thomas-Fermi approximation) \cite{DBS78,butts,Castin}. For example, the bulk density for arbitrary confining potential $V({\bf x})$ is given, within LDA, by  
\begin{eqnarray}\label{eq:rho2}
\rho_N({\bf x}) \approx \frac{1}{N}\left(\frac{m}{2 \pi \hbar^2}\right)^{d/2} \frac{[\mu - V({\bf x})]^{d/2}}{\Gamma(d/2+1)} \;,
\end{eqnarray} 
where $\mu$ is the Fermi energy. This generalizes the semi-circular law (\ref{wigner}) in higher dimensions. Within LDA, one can even estimate accurately the two-point density-density correlation function within the bulk [see Eq.~(\ref{eq:kernel_bulk2})]. However, near the edges (i.e., near the classical turning points where $V({\bf x}) = \mu$), LDA breaks down and the problem becomes
more complex. Several studies have estimated the density and the correlation functions near the edge based on the extension of the semi-classical expansion beyond the classically allowed regime \cite{DBS78, DSB80,BDB84,KM98,Ribeiro2015}. Notably, Kohn and Mattsson introduced an approximate method to calculate the density near the edge in $d=3$ \cite{KM98} -- they called it the {\it edge electron gas}. They provide a closed expression for the edge density in $d=3$, but not for the edge correlation functions. For $d\neq 1,3$, neither the edge density nor the edge correlation function has been computed. In addition, unlike in $d=1$, a systematic analytical approach, valid both in the bulk as well as at the edges, is presently missing.

\begin{figure}[t]
\includegraphics[width=0.9\linewidth]{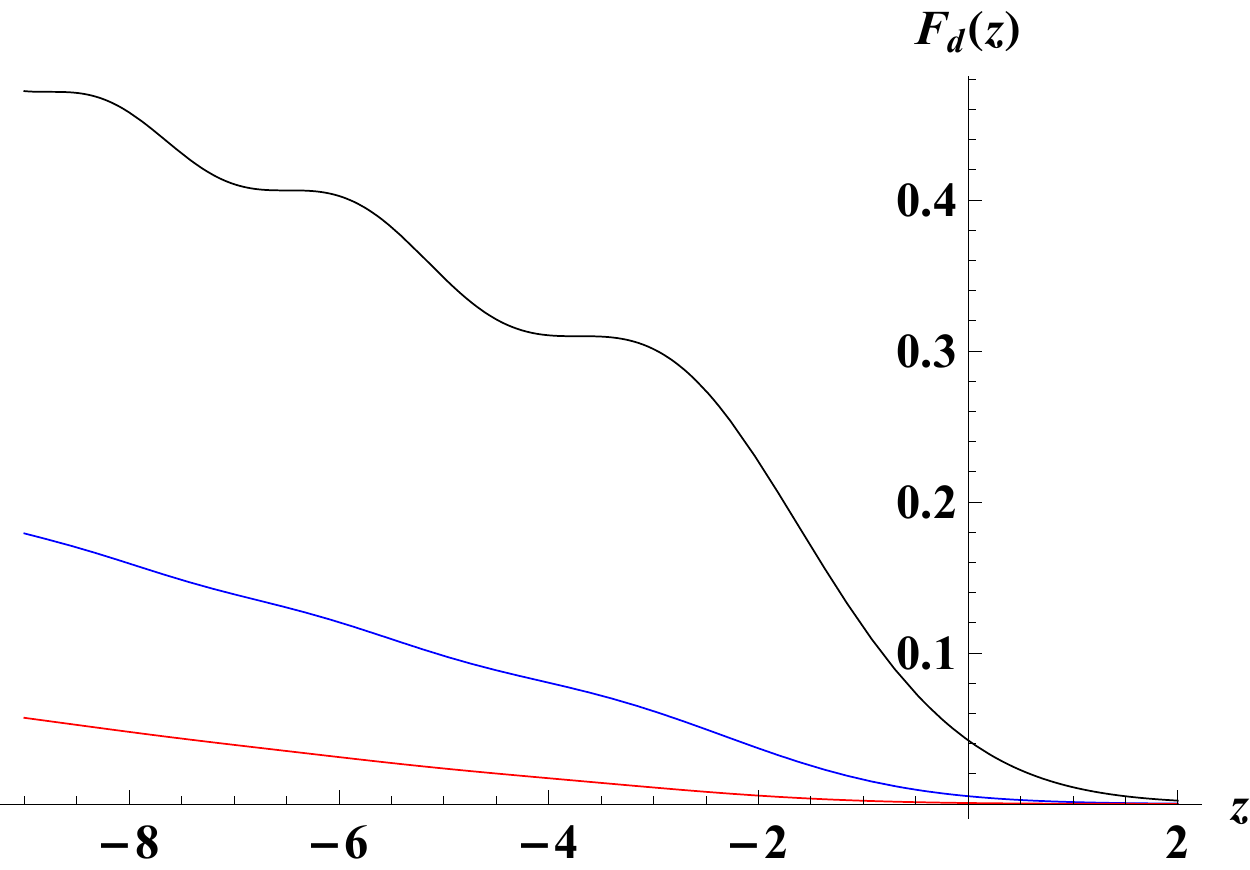}
\caption{(Color online) Plot of the scaling functions $F_d(z)$ in Eq. (\ref{Fd_explicit}) for $d=1,2,3$ (top to bottom) for the density near the edge. The oscillatory structure of the scaling function becomes less pronounced as the dimension $d$ increases.}
\label{edgedensityfd}
\end{figure}

In this Letter, we show that, in any dimension $d$, the $n$-point correlation functions in the ground state have a determinantal structure: it can be expressed as an $n \times n$ determinant $\det K_{\mu}({\bf x}_i,{\bf x}_j)$ with  $1\leq i,j\leq n$ where $K_{\mu}({\bf x},{\bf y})$ is the kernel, which exhibits universal scaling properties for large $N$ (i.e., valid for a generic class of trap potential). This allows us to treat the bulk as well as the edge within the same unified approach and provides a rigorous basis for the edge electron gas in all dimensions. For simplicity, focusing first on the harmonic potential $V({\bf x}) = \frac{1}{2}m\omega^2 r^2$, where $r = |{\bf x}|$, let us
summarize our main results. First we recover the average density (one-point correlation) of the fermions at zero temperature for large $N$ which is indeed given by Eq. (\ref{eq:rho2}) with $V({\bf x}) = \frac{1}{2}m\omega^2 r^2$ and $\mu \approx \hbar \omega [\Gamma(d+1)\,N]^{1/d}$ is the Fermi energy for large $N$.
%
%
The limiting density has a finite support of radius $r_{\rm edge} = (\sqrt{2}/\alpha) \left[\Gamma(d+1)\right]^{1/(2d)} N^{1/(2d)}$ where $\alpha = \sqrt{m\omega/\hbar}$ (see Fig. \ref{bulkdensityshofig} for $d=2$). Our main result is to show that, in a region of width $w_N = b_d \, N^{-\frac{1}{6d}}$ (where~$b_d=\left[\Gamma(d+1)\right]^{-\frac{1}{6d}}/(\alpha \sqrt{2})$) around the edge, the kernel takes a scaling form:\,$K_{\mu}({\bf x},{\bf y}) \approx \frac{1}{w_N^d} {\cal K}_{\rm edge}\left(\frac{{\bf x} - {\bf r}_{\rm edge}}{w_N},\frac{{\bf y} - {\bf r}_{\rm edge}}{w_N}\right)$ with the scaling function 
\begin{equation}\label{eq:Kedge}
{\cal K}_{\rm edge}({\bf a},{\bf b}) = \int \frac{d^d q}{(2 \pi)^d}  e^{-i {\bf q} \cdot ({\bf a} - {\bf b})  } Ai_1\left(2^{\frac{2}{3}} q^2 + \frac{a_n+b_n}{2^{1/3}}\right)\;,
\end{equation}   
where $a_n = {\bf a} \cdot {\bf r}_{\rm edge}/r_{\rm edge}$ and $b_n = {\bf b} \cdot {\bf r}_{\rm edge}/{r}_{\rm edge}$ are projections of ${\bf a}$ and ${\bf b}$ in the radial direction and $Ai_1(z) = \int_{z}^\infty Ai(u) du$. Furthermore the scaling function ${\cal K}_{\rm edge}({\bf a},{\bf b})$ is universal, i.e. valid for any spherically symmetric potential $V({\bf x})$ \cite{footnote_potentials}. Substituting ${\bf a} = {\bf b}$ in Eq.~(\ref{eq:Kedge}), we also obtain the scaling form of the edge density 
\begin{equation}\label{Fd}
\rho_{\rm edge}({\bf x}) = \frac{1}{N} K_{\mu}({\bf x},Ê{\bf x}) \approx \frac{1}{N} \frac{1}{w_N^d} F_d\left(\frac{r-r_{\rm edge}}{w_N}\right) \;,
\end{equation}
with  
\begin{equation}\label{Fd_explicit}
F_d(z)= {1\over \Gamma({d\over 2}+1)2^{\frac{4d}{3}} \pi^{d\over 2}}\int_0^\infty du\  u^{d\over 2}Ai(u+2^{2/3}\,z) \;,
\end{equation}
a plot of which is shown in Fig.~\ref{edgedensityfd} for $d=1,2,3$. In $d=1$ it reduces to the RMT result $F_1(z) = [Ai'(z)]^2 - z [Ai(z)]^2$ mentioned earlier \cite{BB91,For93}. For $d=3$ this result gives a rigorous proof, from first principles, of  the edge electron gas result obtained in \cite{KM98}.

We start with $N$ spinless free fermions in a $d$-dimensional potential $V({\bf x})$. The single particle eigenfunctions $\psi_{\bf k}({\bf x})$ satisfy the Schr\"odinger equation, $\hat H \psi_{\bf k}({\bf x}) = \epsilon_{\bf k} \psi_{\bf k}({\bf x})$, where $\hat H = - \hbar^2/(2m) \nabla^2 + V({\bf x})$ is the Hamiltonian and the energy eigenvalues $\epsilon_{\bf k}$ are labelled by $d$ quantum numbers denoted by ${\bf k}$.  
At zero temperature, the ground state many-body wavefunction can be expressed as an $N \times N$ Slater determinant, $\Psi_0({\bf x}_1, \cdots, {\bf x}_N) = (1/\sqrt{N!}) \, \det[ \psi_{\bf k}(\bf x_j)]$, constructed from the $N$ single particle wavefunctions with energy up to  
the Fermi level $\mu$ \cite{foot_fermilevel}. For a sufficiently confining potential, $\mu$ generically increases with increasing $N$ \cite{eisler_prl,Castin}. For example, for a $d$-dimensional harmonic oscillator $V({\bf x}) \equiv V(r) = m \omega^2 r^2/2$, $\mu \approx \hbar \omega  [\Gamma(d+1) \, N]^{1/d}$. Using $\det(A) \det(B) = \det(AB)$, the squared many-body wavefunction can then be expressed as a determinant 
\begin{equation}\label{eq:psi_0}
|\Psi_0({\bf x}_1, \cdots, {\bf x}_N)|^2=\frac{1}{N!}\det_{1\leq i,j \leq N} K_\mu({\bf x}_i,{\bf x}_j)
\end{equation} 
where the kernel $K_\mu({\bf x},{\bf y})$ is given by
\begin{equation}\label{eq:def_kernel}
K_\mu({\bf x},{\bf y}) =\sum_{\bf k} \theta(\mu-\epsilon_{\bf k}) \psi_{\bf k}^*({\bf x})\psi_{\bf k}( {\bf y}) \;.
\end{equation}
Here $\theta(x)$ is the Heaviside theta function. As in $d=1$ [see Eq. (\ref{rm1})], the squared wavefunction in (\ref{eq:psi_0}) can be interpreted as the joint probability density of $N$ points in a $d$-dimensional space. By integrating out the $N-n$ coordinates of $|\Psi_0({\bf x}_1, \cdots, {\bf x}_N)|^2$ in Eq. (\ref{eq:psi_0}), one can show (see Supp. Mat.~\cite{supp_mat}) that the $n$-point correlation function can be expressed as an $n \times n$ determinant whose entries are given precisely by the kernel $K_\mu({\bf x},{\bf y})$ in Eq.~(\ref{eq:def_kernel}). In particular, for $n=1$, the density $\rho_N({\bf x})$ is given by 
\begin{eqnarray}\label{eq:rel_density_kernel}
\rho_N({\bf x}) = \frac{1}{N} K_\mu({\bf x},{\bf x}) \;.
\end{eqnarray}
Thus, the knowledge of the kernel provides a complete description of the statistical properties of the ground state. 

To compute the kernel, we first recall a well known relation between the kernel and the propagator of the single particle quantum problem \cite{book_dft}. Taking derivative of Eq. (\ref{eq:def_kernel}) with respect to (w.r.t.) $\mu$ and performing a Laplace transform w.r.t. $\mu$ of the resulting relation, one finds
\begin{eqnarray}\label{eq:laplace1}
\hspace*{-0.5cm}\int_0^\infty \frac{\partial K_{\mu}({\bf x}, {\bf y})}{\partial \mu} \, e^{-\mu \frac{t}{\hbar}} \, d\mu = \sum_{\bf k} e^{-t \frac{\epsilon_{\bf k}}{\hbar}} \psi_{\bf k}^*({\bf x})  \psi_{\bf k}({\bf y})  \;.
\end{eqnarray} 
The right hand side of (\ref{eq:laplace1}) is simply the single particle propagator $G({\bf x},{\bf y};t)=\langle {\bf x}|e^{- \frac{t}{\hbar}\hat H }|{\bf y}\rangle$ in imaginary time. Integrating the left hand side of Eq. (\ref{eq:laplace1}) by parts and inverting the Laplace transform, using Bromwich inversion formula, gives (see  \cite{book_dft,BBD85} and Refs. therein)
 \begin{equation}\label{eq:laplace2}
K_\mu({\bf x},{\bf y})= \int_\Gamma {dt\over 2\pi i} \,\frac{1}{t} \exp\left({\mu\over \hbar} t\right)G({\bf x},{\bf y};t)\;,
\end{equation}
where $\Gamma$ denotes the Bromwich integration contour. While Eq. (\ref{eq:laplace2}) is general and holds for arbitrary potential $V({\bf x})$, calculating the propagator $G({\bf x},{\bf y};t)$ explicitly is hard for general $V({\bf x})$. Below, we first focus on the specific case of a harmonic oscillator for which $G$ is known explicitly~\cite{feynman_hibbs}
\begin{eqnarray}\label{eq:propag_hamonic}
G({\bf x},{\bf y};t) = \left(\frac{\alpha^2}{2 \pi \sinh{(\omega \, t)}}\right)^{d/2} e^{-\frac{\alpha^2}{2 \sinh{(\omega\,t)}}  Q({\bf x}, {\bf y};t)}
\end{eqnarray}  
where $Q({\bf x}, {\bf y};t) = ({\bf x}^2 + {\bf y}^2) \cosh{(\omega \, t) - 2 \, {\bf x}\cdot {\bf y}}$. General potentials $V({\bf x})$ will be considered later. 

To extract the large $N$ behavior of the kernel from Eq. (\ref{eq:laplace2}), we need to perform a {\it small $t$} expansion of the propagator   
$G({\bf x},{\bf y};t)$ since $\mu \sim N^{1/d}$ is also large. Similar short time expansions have been carried out previously, especially in the context of nuclear \cite{DBS78,Fujiwara82} as well as chemical \cite{MM88} physics. In this paper, we require such an expansion in general $d$ as detailed in the supplementary material \cite{supp_mat}. This expansion is however different from the Wigner-Kirkwood semi-classical expansion in powers of $\hbar$ \cite{Fujiwara82}.

{\it Global density}. We first evaluate the global density $\rho_N({\bf x})$ in Eq. (\ref{eq:rel_density_kernel}) by putting ${\bf x}= {\bf y}$ in Eq.~(\ref{eq:laplace2}). The dominant contribution to the Bromwich integral in Eq. (\ref{eq:laplace2}) with ${\bf x}= {\bf y}$ comes from the small $t$ region. Expanding the propagator to leading order for small $t$, the integral can be done explicitly to give the result in Eq.~(\ref{eq:rho2}) with $V({\bf x}) = \frac{1}{2}m\omega^2 r^2$ and $\mu \approx \hbar \omega [\Gamma(d+1)\,N]^{1/d}$. The normalization condition $\int d{\bf x} \, \rho_N({\bf x}) = 1$ fixes the Fermi energy $\mu \approx \hbar \omega [\Gamma(d+1)\,N]^{1/d}$. The density thus has a radially symmetric finite support that vanishes at the edge as $\sim (r_{\rm edge} - r)^{d/2}$ (see Fig. \ref{bulkdensityshofig}), where $r_{\rm edge} = \sqrt{2 \mu/(m \omega^2)} \sim N^{1/(2d)}$. Since $N$ particles are packed within a volume of radius $r_{\rm edge} \sim N^{1/(2d)}$, the typical inter-particle distance $\ell_{\rm typ}$ can be estimated very simply: $N \ell_{\rm typ}^d \sim r_{\rm edge}^d$, implying $\ell_{\rm typ} \sim N^{-1/(2d)}$. 

{\it Edge density}. We next investigate the density near $r_{\rm edge}$ for finite but large $N$. To derive the asymptotic edge behavior, we again start with the propagator $G({\bf x},{\bf x};t)$, but now we set $r = |{\bf x}| = r_{\rm edge} + z\, b_d \, N^{-\phi}$ where $\phi$ is yet to be determined and $b_d=\left[\Gamma(1+d)\right]^{-\frac{1}{6d}}/(\alpha \sqrt{2})$. Expanding the propagator for small $t$ and keeping terms up to order ${\cal O}(t^3)$, we find that $\phi = 1/(6d)$ in order that the two leading terms scale in the same way for large $N$ with $z$ fixed (see Supp. Mat.~\cite{supp_mat}). Subsequently, evaluating the kernel in Eq. (\ref{eq:laplace2}) and using Eq.~(\ref{eq:rel_density_kernel}), upon identifying $w_N = b_d\,N^{-1/(6d)}$, 
the edge density satisfies the scaling form in Eq. (\ref{Fd}) where the scaling function is given by
\begin{eqnarray}\label{eq:Fd_laplace}
F_d(z) = (4\pi)^{-d/2} \int_{\Gamma} \frac{d \tau}{2\pi i} \, \frac{1}{\tau^{d/2+1}} \, e^{- \tau\,z + {\tau^3}/{12}} \;.
\end{eqnarray}
Using the integral representation of the Airy function, $Ai(z) = 1/({2\pi i})\int_{\Gamma} {d\tau}\, e^{-\tau z + \tau^3/3}$, 
the integral in Eq.~(\ref{eq:Fd_laplace}) reduces to the expression announced in Eq.~(\ref{Fd_explicit}) (see Supp. Mat.~\cite{supp_mat}). This thus generalizes to arbitrary $d$ the $1d$ result, $F_1(z) = [Ai'(z)]^2 - z [Ai(z)]^2$, obtained from RMT \cite{BB91,For93}. The asymptotic behaviors of $F_d(z)$ can be computed explicitly (see Supp. Mat.~\cite{supp_mat}) with the result
\begin{eqnarray}
F_d(z) &\approx & (8\pi)^{-\frac{d+1}{2}}\, z^{-\frac{d+3}{4}}\, 
e^{-\frac{4}{3}\, 
z^{3/2}}\;{\rm as}\; z\to \infty \label{asymp_plus} \\
&\approx & \frac{(4\pi)^{-\frac{d}{2}}}{\Gamma(d/2+1)}\, |z|^{\frac{d}{2}}
\quad {\rm as}\quad z\to -\infty \;.
\label{asymp_minus}
\end{eqnarray}
One can show that when $z \to -\infty$, i.e., when $r \ll r_{\rm edge}$, the asymptotic behavior in Eq.~(\ref{asymp_minus}) matches smoothly with the bulk density in Eq.~(\ref{eq:rho2}) with $V({\bf x}) = \frac{1}{2}m\omega^2 r^2$ and $\mu \approx \hbar \omega [\Gamma(d+1)\,N]^{1/d}$ .    

{\it Bulk kernel}. We next consider the large $N$ scaling behavior of the kernel 
$K_\mu({\bf x}, {\bf y})$ in Eq.~(\ref{eq:laplace2}) where the two points ${\bf x}$ and ${\bf y}$ are both far from the edge, while their relative separation $|{\bf x} - {\bf y}|$ is on the scale of the inter-particle distance $\ell_{\rm typ}\sim N^{-1/(2d)}$ . We start from Eq.~(\ref{eq:laplace2}) and in Eq. (\ref{eq:propag_hamonic}) we rewrite 
$Q({\bf x}, {\bf y};t) = ({\bf x} - {\bf y})^2 + ({\bf x}^2 + {\bf y}^2) (\cosh(\omega \,t)-1)$. Expanding the propagator for small $t$ to leading order we obtain
\begin{equation}\label{eq:kernel_bulk1}
K_{\mu}({\bf x},{\bf y}) \approx \left(\frac{\alpha^2}{2 \pi \omega}\right)^{\frac{d}{2}} \int_{\Gamma} \frac{dt}{2\pi i} \, \frac{1}{t^{\frac{d}{2}+1}} \, e^{\frac{(\mu - V(|{\bf x}|)) t}{\hbar} - \frac{\alpha^2 ({\bf x} - {\bf y})^2}{2 \omega t}} \;
\end{equation}  
where $V(|{\bf x}|) = V(r) = m\omega^2 r^2/2$. Fortunately, this integral can be done exactly (see Supp. Mat.~\cite{supp_mat}). We find that the bulk kernel has the scaling form, $K_{\mu}({\bf x},{\bf y}) \approx \ell^{-d} {\cal K}_{\rm bulk}(|{\bf x}-{\bf y}|/\ell)$, where $\ell = [N \rho_N({\bf x})\gamma_d]^{-1/d}$ is the typical separation in the bulk and $\gamma_d =  \pi^{d/2}[\Gamma(d/2+1)]$. 
The bulk scaling function is 
given explicitly by
\begin{eqnarray}\label{eq:kernel_bulk2}
{\cal K}_{\rm bulk}(x) = \frac{J_{d/2}(2 x)}{(\pi x)^{d/2}}
\end{eqnarray}
where $J_{d/2}(z)$ is the standard Bessel function of the first kind. In $d=1$, using $J_{1/2}(z) = \sqrt{2/(\pi z)} \sin{z}$, our result in Eq. (\ref{eq:kernel_bulk2}) again reduces to the standard sine-kernel in RMT~\cite{mehta}. The result in Eq. (\ref{eq:kernel_bulk2}) is in full agreement with the heuristic derivation using the LDA~\cite{Castin} (see also Supp. Mat. \cite{supp_mat}). However, the LDA becomes invalid near the edge where the local density is rapidly varying. We will see below that our approach yields exact results even in this edge regime where the LDA fails.


{\it Edge kernel}. Turning to the large $N$ behavior of the kernel $K_\mu({\bf x},{\bf y})$ near the edge, we set ${\bf x} = {\bf r}_{\rm edge} + w_N\, {\bf a}$ and ${\bf y} = {\bf r}_{\rm edge} +w_N\, {\bf b}$, following the scaling of the edge density in Eq.~(\ref{Fd}). Here ${\bf r}_{\rm edge}$ denotes any point on the boundary of the support of the global density with $|{\bf r}_{\rm edge}| = r_{\rm edge} = \sqrt{2\mu/(m \omega^2)} \sim N^{1/(2d)}$. As before, 
the width $w_N = b_d \,N^{-1/(6d)}$ with $b_d = [\Gamma(d+1)]^{-1/(6d)}/(\alpha \sqrt{2})$. Thus ${\bf a}$ and ${\bf b}$ are dimensionless vectors. We substitute these scaling variables ${\bf x}$ and ${\bf y}$ in $Q({\bf x}, {\bf y};t) = ({\bf x} - {\bf y})^2 + ({\bf x}^2 + {\bf y}^2) (\cosh(\omega \,t)-1)$ and expand $Q$ up to order $t^3$ for small $t$. Substituting these results in Eq. (\ref{eq:laplace2}) and  
after a suitable change of variables  (see Supp. Mat.~\cite{supp_mat}), one arrives at the edge kernel 
\begin{equation}\label{eq:edge_kernel}
K_{\mu}({\bf x},{\bf y})\approx \frac{1}{C_d w_N^d} \int_{\Gamma} \frac{d\tau}{2\pi i} \frac{1}{\tau^{\frac{d}{2}+1}} \, e^{-\frac{({\bf a} - {\bf b})^2}{2^{8/3}\tau} - \frac{(a_n + b_n)\tau}{2^{1/3}} + \frac{\tau^3}{3}} \;,
\end{equation}   
with $C_d = (2^{\frac{4}{3}}\sqrt{\pi})^{d}$, and where $a_n = {\bf a} \cdot {\bf r}_{\rm edge}/r_{\rm edge}$ and $b_n = {\bf b} \cdot {\bf r}_{\rm edge}/{r}_{\rm edge}$ are projections of ${\bf a}$ and ${\bf b}$ in the radial direction. One can make a further simplification of Eq.~(\ref{eq:edge_kernel}) by using the integral representation of the diffusive propagator
\begin{eqnarray}\label{eq:diffusion}
\frac{e^{-\frac{({\bf a} - {\bf b})^2}{4 \,D \,\tau}}}{(4\pi D\,\tau)^{\frac{d}{2}}} = \int \frac{d^d q}{(2 \pi)^d} \, e^{- D\,q^2 \tau - i {\bf q}\cdot({\bf a} -{\bf b} ) } \;.
\end{eqnarray}  
We choose $D = 2^{2/3}$ and use this in Eq.~(\ref{eq:edge_kernel}). Using subsequently the integral representation of the Airy function $Ai(z)$ mentioned earlier, we arrive at the main formula given in (\ref{eq:Kedge}) for the scaling behavior of the edge kernel. Putting ${\bf a} = {\bf b}$ in Eq. (\ref{eq:Kedge}), followed by an integration by parts, one can check that ${\cal K}_{\rm edge}({\bf a},{\bf a})$ reduces to $F_d(|{\bf a}|)$ in Eq.~(\ref{Fd_explicit}). Also one can verify, after a few steps of algebra (see Supp. Mat.~\cite{supp_mat}), that in $d=1$ Eq. (\ref{eq:Kedge}) reduces to ${\cal K}_{\rm edge}(a,b) =   K_{\rm Airy}(a,b)= (Ai(a)\,Ai'(b) - Ai'(a)\,Ai(b))/(a-b)$ is the standard Airy kernel \cite{For93,TW}. 

{\it General potential.} One naturally wonders to what extent these results for the harmonic oscillator are universal, i.e., hold for more general potentials $V({\bf x})$. We first note that for general $V({\bf x})$, Eq.~(\ref{eq:laplace2}) still holds, though the Fermi energy $\mu$ and the propagator $G({\bf x}, {\bf y};t)$ depend on $V({\bf x})$. The dependence of the 
Fermi energy $\mu$ on $N$ can be easily estimated for large $N$ using semi-classical approximation \cite{eisler_prl,Castin}. In contrast, $G({\bf x}, {\bf y};t)$ is hard to compute for general $V({\bf x})$. However, as discussed after Eq. (\ref{eq:laplace2}), for large $N$, we only need the small $t$ expansion of $G$ in general dimension $d$ for arbitrary potential $V({\bf x})$. Using the results from \cite{supp_mat}, we find that to leading order in $t$ the global density is given~by Eq. (\ref{eq:rho2}).
%
Evaluating Eq. (\ref{eq:laplace2}) near the edge, as in the harmonic oscillator case, we find that the edge density is again given exactly by Eq. (\ref{Fd}), where only the location of the edge ${\bf r}_{\rm edge}$ and the width $w_N$ depend non-universally on $V({\bf x})$, but the scaling function $F_d(z)$ is universal, i.e., is the same for all spherically symmetric potential $V({\bf x}) = V(r)$ \cite{footnote_potentials}. The bulk kernel is also given as in Eq. (\ref{eq:kernel_bulk2}), the only $V({\bf x})$-dependence appears in the scale factor $\gamma_d$ used in defining the pair of dimensionless vectors ${\bf u}$ and ${\bf v}$. Similarly, we find the edge kernel, appropriately centered and scaled, is also universal (see Supp. Mat. \cite{supp_mat}).

{\it Conclusion.} In this Letter, we have studied the ground state properties of $N$ spinless free fermions in a $d$-dimensional confining trap. We have shown that the $n$-point correlation functions have a determinantal structure for all $d$, with large $N$ scaling forms that are universal, i.e., independent of the details of the trap potential. Our results recover the bulk properties predicted by the heuristic LDA. However, near the edge where this approximation fails, our method predicts 
new universal exact results in all dimensions $d$. In $d=1$, we recover the results from RMT and in $d=3$, our results recover rigorously the edge density result of Kohn and Mattsson \cite{KM98}. Furthermore we provide an explicit expression in all $d$ for the $n$-point correlation functions at the edge. Our results can be extended to finite temperature \cite{us_long} in all dimensions. 
%
%
%
Finally, it would be interesting to see whether one can measure these universal edge scaling functions in experiments on cold atoms.

\acknowledgments{We thank C. Salomon for useful discussions. We acknowledge support from PSL grant ANR-10-IDEX-0001-02-PSL
(PLD) ANR grant 2011-BS04-013-01 WALKMAT and in part by the Indo-French
Centre for the Promotion of Advanced Research under Project 4604-3 (SM and GS).}

\newpage

\onecolumngrid

\begin{center}
{\Large Supplementary Material  \\ 
}
\end{center}

We give the principal details of the calculations described in the manuscript of the Letter.

\section{Determinantal structure for the $n$-point correlation function}

Consider a situation where the joint distribution of $N$ points $|\Psi_0({\bf x}_1, \cdots, {\bf x}_N)|^2$ can be expressed as a determinant 
\begin{equation}\label{eq:psi_0_supp}
|\Psi_0({\bf x}_1, \cdots, {\bf x}_N)|^2=\frac{1}{N!}\det_{1\leq i,j \leq N} K_\mu({\bf x}_i,{\bf x}_j)
\end{equation} 
where the kernel $K_\mu({\bf x},{\bf y})$ has the reproducing property {(which is the case for 
the kernel (8) in the Letter, as it is easy to check)}
\begin{eqnarray}\label{eq:convolution}
\int K_\mu({\bf x},{\bf z})K_\mu({\bf z},{\bf y})\, d{\bf z} = K_\mu({\bf x},{\bf y}) \;.
\end{eqnarray}
The $n$-point correlation function $R_n({\bf x}_1,\cdots, {\bf x}_n)$ is defined as
\begin{eqnarray}\label{eq:def_Rn}
R_n({\bf x}_1,\cdots, {\bf x}_n) = \frac{N!}{(N-n)!} \int d{\bf x}_{n+1} \cdots d {\bf x}_N \, |\Psi_0({\bf x}_1, \cdots, {\bf x}_N)|^2
\end{eqnarray}
obtained by integrating over $N-n$ coordinates and keeping $n$ coordinates $\{{\bf x}_1,\cdots, {\bf x}_n\}$ fixed. If the kernel satisfies the reproducing property in Eq.~(\ref{eq:convolution}), then there is a general theorem \cite{mehta_supp} that states that $R_n$ can be expressed as an $n\times n$ determinant 
\begin{eqnarray}\label{eq:Rn_det}
R_n({\bf x}_1,\cdots, {\bf x}_n) = \det_{1\leq i,j \leq n} K_\mu({\bf x}_i,{\bf x}_j) \;.
\end{eqnarray}

\section{Derivation of the edge density in Eq. (13) in the main text}\label{edge_density}

The single particle propagator for the $d$-dimensional harmonic oscillator is given by \cite{feynman_hibbs_supp}
\begin{eqnarray}\label{eq:propag_hamonic_supp}
G({\bf x},{\bf y};t) = \left(\frac{\alpha^2}{2 \pi \sinh{(\omega \, t)}}\right)^{d/2} \exp{\left[-\frac{\alpha^2}{2 \sinh{(\omega\,t)}}  \left(({\bf x}^2 + {\bf y}^2) \cosh{(\omega \, t) - 2 \, {\bf x}\cdot {\bf y}}\right) \right]}
\end{eqnarray}  
where $\alpha = \sqrt{m\omega/\hbar}$. For the density [see Eqs. (9) and (11) in the main text], we need $G({\bf x},{\bf x};t)$, which is given by
\beq\label{eq:propag_hamonic2}
G({\bf x},{\bf x};t) = \left(\frac{\alpha^2}{2 \pi \sinh{(\omega \, t)}}\right)^{d/2}\exp{\left[-\tanh{\left(\frac{\omega \, t}{2} \right)} \alpha^2 r^2 \right]} \;,
\eeq
where $r= |{\bf x}|$. The kernel $K_\mu({\bf x}, {\bf y})$ is given by Eq. (11) in the main text, which reads
\begin{equation}\label{eq:laplace2_supp}
K_\mu({\bf x},{\bf y})= \int_\Gamma {dt\over 2\pi i} \,\frac{1}{t} \exp\left({\mu\over \hbar} t\right)G({\bf x},{\bf y};t)\;,
\end{equation}
where $\Gamma$ denotes the Bromwich integration contour To evaluate the density $\rho_N({\bf x}) = (1/N) K_\mu({\bf x},{\bf x})$ for large $N$ (equivalently for large $\mu$) using Eq. (\ref{eq:laplace2_supp}), we need the small $t$ expansion of $G({\bf x},{\bf x};t)$ in Eq. (\ref{eq:propag_hamonic2}). For small $t$, 
up to two leading orders, $\tanh{(\omega\,t/2)} \approx \omega\,t/2 - (\omega \, t)^3/24$. To compute the scaling behavior of the edge density, we set $r = r_{\rm edge} +  z \, b_d \, N^{-\phi}$ where $r_{\rm edge} = \sqrt{2 \mu/(m \omega^2)}$, $b_d =  [\Gamma(d+1)]^{-\frac{1}{6d}}/(\alpha \sqrt{2})$ and the exponent $\phi$ is yet to be determined. 
We substitute this in Eq.~(\ref{eq:propag_hamonic2}) and use the resulting expression for $G({\bf x},{\bf x};t)$ in Eq. (\ref{eq:laplace2_supp}).  
Keeping terms up to order ${\cal O}(t^3)$, we get  
\beq\label{eq:propag_hamonic3}
K_\mu({\bf x},{\bf x};t) \approx \left(\frac{\alpha^2}{2\pi \omega}\right)^{d/2} \int_\Gamma \frac{d\,t}{2\pi i} \frac{1}{t^{d/2+1}} e^{\frac{\mu}{\hbar}t}\exp{\left[-\frac{\mu}{\hbar}t -\sqrt{\frac{2\mu}{m}} \alpha^2 \, z\,b_d N^{-\phi} t - \frac{d}{12}\omega^2 \,t^2+ \frac{\mu \, \omega^2}{12 \hbar} t^3\right]} \;.
\eeq 
Note that the order ${\cal O}(t^2)$ term inside the exponential comes from the expansion of the prefactor $[\sinh{(\omega\,t)}]^{-d/2}$. The leading order terms $\mu t/\hbar$ cancel inside the argument of the exponential, leaving only three terms. Using $\mu \approx \hbar \omega [\Gamma(d+1) N]^{1/d}$, we see that the three leading terms inside the exponential scale respectively as, $T_1 \sim N^{1/(2d)-\phi} \,t$, $T_2 \sim t^2$ and $T_3 \sim N^{1/d}\,t^3$. In the large $N$ limit, we keep $T_1 \sim N^{1/(2d)-\phi} \,t \sim \tau$ of order ${\cal O}(1)$. Then the next two terms scale as $T_2 \sim N^{2\phi - 1/d} \, \tau^2$ and $T_3 \sim N^{3\phi - 1/(2d)}\, \tau^3$. We then have two possibilities: (i) choose $\phi = 1/(2d)$ that keeps $T_2 \sim \tau^2 \sim {\cal O}(1)$, but then $T_3\sim N^{1/d} \, \tau^3$ diverges as $N \to \infty$, (ii) choose $\phi = 1/(6d)$ that keeps $T_3 \sim \tau^3 \sim {\cal O}(1)$ and $T_2 \sim N^{-2/(3d)}\, \tau^2$ that vanishes as $N \to \infty$. Clearly the second choice, $\phi = 1/(6d)$, is the correct one to make the scaling consistent. With this choice $\phi = 1/(6d)$, one can also check that terms of order higher than ${\cal O}(t^3)$ vanish as $N \to \infty$.
Hence, rescaling $t = N^{-{1}/{(3d)}}/(\omega [\Gamma(d+1)]^{{1}/{(3d)}}) \, \tau$, one arrives finally at  
\beq\label{eq:defFd}
K_\mu({\bf x},{\bf x}) = \frac{1}{w_N^d} F_d\left(\frac{r-r_{\rm edge}}{w_N} \right)
\eeq 
where $w_N = b_d \, N^{-\frac{1}{6d}}$,~with~$b_d=\left[\Gamma(1+d)\right]^{-\frac{1}{6d}}/(\alpha \sqrt{2})$ and the scaling function $F_d(z)$ is given by 
\begin{eqnarray}\label{eq:Fd_laplace_supp}
F_d(z) = (4\pi)^{-d/2} \int_{\Gamma} \frac{d \tau}{2\pi i} \, \frac{1}{\tau^{d/2+1}} \, e^{- \tau\,z + {\tau^3}/{12}} \;.
\end{eqnarray}
Finally the edge density $\rho_{\rm edge}({\bf x}) = (1/N) K_\mu({\bf x},{\bf x})$ is then given by Eq. (5) in the main text.

\section{Derivation of Eq. (6) in the main text}

Here we show that Eq. (\ref{eq:Fd_laplace_supp}) is identical to the expression given in Eq. (6) in the main text. We first use the identity 
\beq\label{eq:identity}
\frac{1}{\tau^{d/2+1}} = \frac{1}{\Gamma(d/2+1)} \int_0^\infty e^{-\tau x} \, x^{d/2} \, dx
\eeq
in Eq. (\ref{eq:Fd_laplace_supp}) to obtain
\beq\label{eq:identity2}
F_d(z) = \frac{1}{\Gamma(d/2+1) \, (4 \pi)^{d/2}} \int_{0}^\infty dx \, x^{d/2} \int_\Gamma \frac{d\tau}{2\pi i} e^{-\tau(x+z)+ {\tau^3}/{12}} \;.
\eeq
After rescaling $\tau \to 2^{2/3} \tau$ and using the integral representation of the Airy function 
\beq\label{eq:airy_integral}
Ai(z) = \int_{\Gamma} \frac{d\tau}{2\pi i}\, e^{-\tau z + \tau^3/3}
\eeq
we get 
\beq\label{eq:identity3}
F_d(z) = \frac{2^{2/3}}{\Gamma(d/2+1) \, (4 \pi)^{d/2}} \int_0^\infty dx \, x^{d/2} \, Ai\left(2^{2/3}(x+z) \right) \;.
\eeq
With a further rescaling {$x \to 2^{-2/3} x$}, $F_d(z)$ in Eq. (\ref{eq:identity3}) reduces to the form given in Eq. (6) in the main text which reads
\begin{equation}\label{Fd_explicit_supp}
F_d(z)= {1\over \Gamma({d\over 2}+1)2^{\frac{4d}{3}} \pi^{d\over 2}}\int_0^\infty du\  u^{d\over 2}Ai(u+2^{2/3}\,z) \;.
\end{equation}
In $d=1$, the integral can be performed exactly. We start with the identity \cite{vallee_supp}
\begin{eqnarray}\label{eq:def_Iz}
\int_0^\infty Ai(z+u) \frac{du}{\sqrt{u}} = 2^{2/3} \pi Ai^2\left(\frac{z}{2^{2/3}}\right) \equiv I(z) \;
\end{eqnarray}
and differentiate it twice with respect to $z$. Using the Airy differential equation $Ai''(z) = z Ai(z)$, one obtains 
\begin{eqnarray}\label{final_identity}
\int_0^\infty du \, \sqrt{u} Ai(z+u)  = I''(z) - z I(z) = \pi 2^{1/3} \left(\left[Ai'\left(\frac{z}{2^{2/3}} \right)\right]^2  - \frac{z}{2^{2/3} }Ai^2\left(\frac{z}{2^{2/3}} \right) \right)
\end{eqnarray} 
It then follows from Eq. (\ref{Fd_explicit_supp}), upon setting $d=1$, that
\begin{equation}
F_1(z) = Ai'^2(z) - z Ai^2(z) \;,
\end{equation}
thus recovering the well known RMT result \cite{BB91_supp,For93_supp}. {One obtains similar quadratic
forms in $Ai(z)$ and $Ai'(z)$ with polynomial coefficients in $z$ in
any odd space dimension by repeated application of the Airy operator $(\partial_z^2 - z)$ on $I(z)$.
For instance in $d=3$:
\begin{eqnarray}
F_3(z) = \frac{1}{12 \pi} ( 2 z^2 Ai(z)^2 - Ai(z) Ai'(z) - 2 z Ai'(z)^2 )
\end{eqnarray}
In $d=2$ one can use Airy equation and find 
\begin{eqnarray}
F_2(z) = \frac{1}{2^{\frac{8}{3}} \pi} ( - Ai'(2^{\frac{2}{3}} z) - 2^{\frac{2}{3}} z Ai_1(2^{\frac{2}{3}} z)) 
\end{eqnarray}
where $Ai_1(z)=\int_z^\infty dx Ai(x)$ also appears in Eq. (20) and can be expressed 
in terms of hypergeometric functions.
}

\vspace*{0.5cm}

{\it Asymptotic behaviors of $F_d(z)$.} We first consider the $z\to +\infty$ limit. In this limit the Airy function has the leading asymptotic behavior \cite{vallee_supp}
\beq\label{airy_largez}
Ai(z) \sim \frac{1}{2\sqrt{\pi}} \, z^{-1/4} \, \exp{\left(-\frac{2}{3} z^{3/2}\right)} \;.
\eeq
Substituting this asymptotic behavior in Eq. (\ref{Fd_explicit_supp}), expanding for large $z$, one gets to leading order  
\beq
F_d(z) \approx  (8\pi)^{-\frac{d+1}{2}}\, z^{-\frac{d+3}{4}}\, 
\exp{\left(-\frac{4}{3}\, 
z^{3/2}\right)}\;{\rm as}\; z\to \infty \label{asymp_plus_supp} \;.
\eeq
For the other side $z \to -\infty$, it is more convenient to use the representation in Eq. (\ref{eq:Fd_laplace_supp}). We set $z = -|z|$ and scale $\tau\,|z| = t$. This makes the order $\tau^3$ term to be $|z|^{-3} \, t^3/12$ which can then be dropped for large $|z|$. The resulting Bromwich contour integral  can be easily evaluated to give the leading asymptotic behavior
\beq
F_d(z)\approx \frac{(4\pi)^{-\frac{d}{2}}}{\Gamma(d/2+1)}\, |z|^{\frac{d}{2}}
\quad {\rm as}\quad z\to -\infty \;.
\label{asymp_minus_supp}
\eeq

\section{Derivation of the bulk kernel}

In order to analyze the bulk kernel, it turns out to be convenient to rewrite the 
propagator $G({\bf x},{\bf y};t)$ given in Eq. (\ref{eq:propag_hamonic_supp}) in a slightly different form
\beq\label{propag_bulk}
G({\bf x},{\bf y};t) = \left(\frac{\alpha^2}{2 \pi \sinh{(\omega \, t)}}\right)^{d/2} \exp{\left[-\frac{\alpha^2}{2 \sinh{(\omega\,t)}}  \left( ({\bf x}- {\bf y})^2 + ({\bf x}^2 + {\bf y}^2)(\cosh(\omega\,t)-1) \right) \right]} \;.
\eeq
Substituting this in Eq. (\ref{eq:laplace2_supp}) and expanding the propagator to leading order in small $t$ leads to Eq. (16) in the main text that reads
\begin{equation}\label{eq:kernel_bulk1_supp}
K_{\mu}({\bf x},{\bf y}) \approx \left(\frac{\alpha^2}{2 \pi \omega}\right)^{\frac{d}{2}} \int_{\Gamma} \frac{dt}{2\pi i} \, \frac{1}{t^{\frac{d}{2}+1}} \, e^{\frac{(\mu - V(|{\bf x}|)) t}{\hbar} - \frac{\alpha^2 ({\bf x} - {\bf y})^2}{2 \omega t}} \;,
\end{equation}  
where $V(|{\bf x}|)=V(r) = m\omega^2 r^2/2$. We then use the following integral representation \cite{Grad_supp}
\beq\label{eq:bessel}
\int_\Gamma \frac{dt}{2\pi i} \frac{1}{t^{d/2+1}} e^{z\,t - a/t} = \left(\frac{z}{a}\right)^{d/4} \, J_{d/2}\left(2\sqrt{a\,z}\right),
\eeq 
where $J_\nu(x)$ is the standard Bessel function of the first kind of index $\nu$. We use this identity to evaluate the kernel in Eq. (\ref{eq:kernel_bulk1_supp}). The result gets simplified upon replacing $\mu - V(r)$ in terms of the density using 
\beq\label{eq:density}
\rho_N({\bf x}) \approx \frac{1}{N}\left(\frac{m}{2 \pi \hbar^2}\right)^{d/2} \frac{[\mu - \frac{1}{2}m\omega^2 r^2]^{d/2}}{\Gamma(d/2+1)} \;.
\eeq
The result can be cast in a simple scaling form 
\begin{eqnarray}\label{scaling_bulk}
K_{\mu}({\bf x},{\bf y}) \approx \ell^{-d} {\cal K}_{\rm bulk}(|{\bf x}-{\bf y}|/\ell)
\end{eqnarray}
where $\ell = [N \rho_N({\bf x})\gamma_d]^{-1/d}$ is the typical separation in the bulk and 
{$\gamma_d =  \pi^{d/2} [\Gamma(d/2+1)]$}. The bulk scaling function is 
given explicitly by
\begin{eqnarray}\label{eq:kernel_bulk2_supp}
{\cal K}_{\rm bulk}(x) = \frac{J_{d/2}(2 x)}{(\pi x)^{d/2}}
\end{eqnarray}
as in Eq. (17) in the main text, {with ${\cal K}_{\rm bulk}(0)= 1/\gamma_d$. In $d=1$, using $J_{1/2}(z) = \sqrt{2/(\pi z)} \sin{z}$, one recovers the standard sine-kernel ${\cal K}_{\rm bulk}(x) = \frac{\sin(2 x)}{\pi x}$ in RMT~\cite{mehta_supp}.

{It is important to note that this kernel is the Fourier transform of a
Fermi-step function, using the formula:
\begin{eqnarray}
\int_{|{\bf k}|<k_f} \frac{d^d k}{(2 \pi)^d} e^{i \bf k \cdot \bf x} =  \left(\frac{k_f}{2 \pi |{\bf x}|}\right)^{d/2} J_{d/2}(k_f |{\bf x}|) \;,
\end{eqnarray}
which can be verified in a straightforward way. To recover our result one must
thus choose the local Fermi momentum as $k_f = k_f({\bf x})= 2/\ell =2 (N \rho_N({\bf x}) \gamma_d)^{1/d}$.
This value is exactly consistent with the one obtained by the counting of states for a uniform system of density
$\rho_N({\bf x})$, by setting $N \, \rho_N({\bf x}) = \int_{|{\bf k}|<k_f}  \frac{d^d k}{(2 \pi)^d}  = k_f^d/(4 \pi)^{d/2} \Gamma(1+d/2)$. Thus, to describe correlations on scale $\ell$ in the bulk,
one can approximate locally the system by free
fermions without any external potential, but at a fixed density $\rho_N({\bf x})$,
assumed to be slowly varying on that scale.} This then provides a more rigorous derivation of the results, within the bulk, obtained from the heuristic local density approximation.

\section{Derivation of the edge kernel}\label{edge_kernel}

In order to analyze the edge kernel we start with the propagator $G({\bf x},{\bf y};t)$ given in Eq. (\ref{propag_bulk}).  
Near the edge, we set ${\bf x} = {\bf r}_{\rm edge} + w_N\, {\bf a}$ and ${\bf y} = {\bf r}_{\rm edge} + w_N\, {\bf b}$. Here ${\bf r}_{\rm edge}$ denotes any point on the boundary of the support of the global density with $|{\bf r}_{\rm edge}| = r_{\rm edge} = \sqrt{2\mu/(m \omega^2)} \sim \sqrt{2}/\alpha [\Gamma(d+1)]^{1/(2d)}\, N^{1/(2d)}$. As in section II for the edge density, we choose $w_N = b_d \, N^{-1/(6d)}$ with $b_d = [\Gamma(d+1)]^{-1/(6d)}/(\alpha \sqrt{2})$. Thus $w_N$ denotes the width of the edge regime. The vectors ${\bf a}$ and ${\bf b}$ are thus dimensionless. As in the case of the edge density in section \ref{edge_density} above, we expand the propagator for small $t$ up to order ${\cal O}(t^3)$. Note that, for the edge kernel, we need to expand the propagator up to terms  of order ${\cal O}(t^3)$ (as opposed to the bulk kernel where it is sufficient to keep terms up to order ${\cal O}(t)$). As in section \ref{edge_density}, in the scaling limit, the term of order ${\cal O}(t^2)$ drops out for large $N$, while the terms of order ${\cal O}(t)$ and $t^3$ scale in the same way. Proceeding further as in section \ref{edge_density}, it is then straightforward to arrive at Eq. (18) in the main text that reads
\begin{equation}\label{eq:edge_kernel_supp}
K_{\mu}({\bf x},{\bf y})\approx \frac{1}{C_d w_N^d} \int_{\Gamma} \frac{d\tau}{2\pi i} \frac{1}{\tau^{\frac{d}{2}+1}} \, e^{-\frac{({\bf a} - {\bf b})^2}{2^{8/3}\tau} - \frac{(a_n + b_n)\tau}{2^{1/3}} + \frac{\tau^3}{3}} \;,
\end{equation}   
with $C_d = (2^{\frac{4}{3}}\sqrt{\pi})^{d}$ and where $a_n = {\bf a} \cdot {\bf r}_{\rm edge}/r_{\rm edge}$ and $b_n = {\bf b} \cdot {\bf r}_{\rm edge}/{r}_{\rm edge}$ are projections of ${\bf a}$ and ${\bf b}$ in the radial direction. One can make a further simplification of Eq.~(\ref{eq:edge_kernel_supp}) by using the integral representation of the diffusive propagator
\begin{eqnarray}\label{eq:diffusion_supp}
\frac{e^{-\frac{({\bf a} - {\bf b})^2}{4 \,D \,\tau}}}{(4\pi D\,\tau)^{\frac{d}{2}}} = \int \frac{d^d q}{(2 \pi)^d} \, e^{- D\,q^2 \tau - i {\bf q}\cdot({\bf a} -{\bf b} ) } \;.
\end{eqnarray}  
We choose $D = 2^{2/3}$ and use this in Eq.~(\ref{eq:edge_kernel_supp}). This gives the scaling behavior of the edge kernel, 
\begin{eqnarray}\label{eq:def_edge_kernel}
K_{\mu}({\bf x},{\bf y}) \approx \frac{1}{w_N^d} {\cal K}_{\rm edge}\left(\frac{{\bf x} - {\bf r}_{\rm edge}}{w_N},\frac{{\bf y} - {\bf r}_{\rm edge}}{w_N}\right) \;,
\end{eqnarray}
where the scaling function is given explicitly by
\begin{eqnarray}\label{explicit_edge1}
{\cal K}_{\rm edge}({\bf a},{\bf b}) = \int \frac{d^d q}{(2\pi)^d} e^{-i {\bf q}\cdot ({\bf a}- {\bf b})}  \int_\Gamma \frac{d\tau}{2\pi i} \frac{1}{\tau} e^{-(2^{2/3} q^2 + 2^{-1/3}(a_n+b_n))\tau + \tau^3/3} 
\end{eqnarray}
Defining $Ai_1(z)$ as
\beq\label{eq:Wz}
Ai_1(z) = \int_\Gamma \frac{d \tau}{2\pi i} \frac{1}{\tau} e^{-z \tau + \tau^3/3} \;,
\eeq
we arrive at
\begin{equation}\label{eq:edge_kernel3}
{\cal K}_{\rm edge}({\bf a},{\bf b}) = \int \frac{d^d q}{(2 \pi)^d}  e^{-i {\bf q} \cdot ({\bf a} - {\bf b})  } Ai_1\left(2^{\frac{2}{3}} q^2 + \frac{a_n+b_n}{2^{1/3}}\right)\;.
\end{equation}   
The function $Ai_1(z)$ defined in Eq. (\ref{eq:Wz}) can be actually expressed in terms of the Airy function itself. Indeed, differentiating Eq. (\ref{eq:Wz}) with respect to $z$    
and using the integral representation of the Airy function in Eq. (\ref{eq:airy_integral}), 
one finds that $Ai_1'(z) = -Ai(z)$. Reintegrating with respect to $z$ and using that $Ai_1(z \to \infty) \to 0$, we get
\beq\label{eq:Wz_Airy}
Ai_1(z) = \int_z^\infty  Ai(u) \, du \;.
\eeq

\section{Reduction to the standard Airy kernel in $d=1$}
 
Putting $d=1$ in Eq. (\ref{eq:edge_kernel3}) we get
\beq\label{eq:airy_kernel1}
{\cal K}_{\rm edge}(a,b) =  \int_{-\infty}^\infty \frac{dq}{2\pi} e^{i q (a-b)} \int_{2^{2/3}q^2+2^{-1/3}(a+b)}^\infty Ai(z) \, dz \;. 
\eeq
Making a shift $z = 2^{2/3}q^2 + 2^{-1/3}(a+b) +u$ gives
\beq\label{eq:airy_kernel2}
{\cal K}_{\rm edge}(a,b) =  \int_{-\infty}^\infty \frac{dq}{2\pi} e^{i q (a-b)} \int_{0}^\infty Ai(u+2^{2/3}q^2 + 2^{-1/3}(a+b)) \, du \;. 
\eeq
We next use a non-trivial identity involving Airy functions \cite{vallee_supp}
\beq\label{eq:identity_airy}
\int_{-\infty}^\infty \frac{dq}{2 \pi}\, e^{-iq\,(v-v')} \, Ai(2^{2/3}q^2 + 2^{-1/3}(v+v')) = 2^{-\frac{2}{3}} Ai(v)Ai(v') \;.
\eeq 
Choosing $v = a + 2^{-2/3} u$ and $v' = b + 2^{-2/3} u$, substituting this identity in Eq. (\ref{eq:airy_kernel2}) and rescaling $u \to 2^{-2/3}u$ gives
\beq\label{eq:airy_kernel3}
{\cal K}_{\rm edge}(a,b) =  \int_0^\infty du \, Ai(a+u) \, Ai(b+u) \;.
\eeq
Since $Ai(z)$ satisfies the differential equation $Ai''(z)- z Ai(z)=0$ we replace $Ai(z)$ by $Ai''(z)/z$ in Eq. (\ref{eq:airy_kernel3}). Next we use the identity 
\beq
\frac{1}{(u+a)(u+b)} = \frac{1}{b-a} \left[\frac{1}{u+a} - \frac{1}{u+b} \right] 
\eeq
and integrate by parts. This then reduces Eq. (\ref{eq:airy_kernel3}) to the standard Airy kernel form 
\beq
{\cal K}_{\rm edge}(a,b) = K_{\rm Airy}(a,b) = (Ai(a)\,Ai'(b) - Ai'(a)\,Ai(b))/(a-b)  \;.
\eeq

\section{Kernel for general potential}

We start from Eq. (\ref{eq:laplace2_supp}) [Eq. (11) in the main text] for the kernel where $G({\bf x},{\bf y};t)=\langle {\bf x}|e^{- \frac{t}{\hbar}\hat H }|{\bf y}\rangle$ with the single particle Hamiltonian $\hat H = - \hbar^2/(2m) \nabla^2 + V({\bf x})$. Expanding in the energy eigenbasis one gets 
\beq\label{eq:genV1}
G({\bf x},{\bf y};t) = \sum_{E} \Psi_E({\bf x}) \Psi_E({\bf y}) e^{-E\,t/\hbar} \;,
\eeq
where $\Psi_E({\bf x})$ satisfies the Schr\"odinger equation, $-\frac{\hbar^2}{2m} \nabla^2 \psi_E + V\, \Psi_E = E\, \Psi_E$. It is then easy to verify that $G$ satisfies the Feynmac-Kac equation
\beq\label{eq:genV2}
\partial_t G = \frac{\hbar}{2m} \nabla^2 G - \frac{1}{\hbar} V\, G \;.
\eeq
To evaluate the kernel in Eq. (\ref{eq:laplace2_supp}) for large $N$ (and hence for large $\mu$), we need the short time expansion of the propagator $G({\bf x},{\bf y};t)$ for general $V({\bf x})$. The Feynmac-Kac equation (\ref{eq:genV2}) is a good starting point for a perturbative small $t$ expansion. We have computed explicitly the terms up to order ${\cal O}(t^3)$ using the perturbation theory of the Feynman-Kac equation \cite{us_long_supp}. Omitting details \cite{us_long_supp}, we just quote the main results here up to order ${\cal O}(t^3)$:
\begin{eqnarray}\label{eq:short_time}
&&G({\bf x},{\bf y};t) \sim \left(m\over 2\pi \hbar t\right)^{d\over 2}\exp\left[-{m\over 2\hbar t}({
\bf x}-{\bf y})^2\right]\exp\left[-{t\over \hbar}S_1({\bf x},{\bf y})  -{t^2\over 2m}S_2({\bf x},{\bf y}) + {t^3\over 2m\hbar}S_3({\bf x},{\bf y})\right],
\end{eqnarray}
where 
\begin{eqnarray}
S_1({\bf x},{\bf y}) &=& \int_0^1du\   V({\bf x}+ ({\bf y}-{\bf x})\,u), \label{S1}\\
S_2({\bf x},{\bf y}) &=&\int_0^1du\  u(1-u)  (\nabla^2V)({\bf x}+ ({\bf y}-{\bf x}) \, u),\label{S2}\\
S_3({\bf x},{\bf y}) &=& \int_0^1du\int_0^1dv\ \left[
{\rm min}(u,v)-uv\right] (\nabla V)({\bf x}+ ({\bf y}-{\bf x}) \, u)\cdot(\nabla V)({\bf x}+ ({\bf y}-{\bf x})\, v)\nonumber \\
&-&{\hbar^2\over 4m} \int_0^1du \ u^2(1-u)^2 (\nabla^4 V)({\bf x}+ ({\bf y}-{\bf x})\, u) \;, \label{S3}
\end{eqnarray}
where $(\nabla^2 V)({\bf x}+ ({\bf y}-{\bf x})\, u)$ denotes the Laplacian evaluated at $({\bf x}+ ({\bf y}-{\bf x})\, u)$ and similarly for $(\nabla V)({\bf x}+ ({\bf y}-{\bf x})\,u)$ and $(\nabla^4 V)({\bf x}+ ({\bf y}-{\bf x})\, u)$. While the expressions for $S_1$ and $S_2$ have appeared in the literature before 
obtained by analyzing a discretized version of the path integral \cite{MM88_supp}, the expression for $S_3$ has not been computed before, to the best of our knowledge. As a simple illustration, one can verify that for the $1d$ harmonic oscillator, $V(x) = \frac{1}{2}m \omega^2 x^2$ one gets (setting for simplicity $\hbar = \omega = m = 1$):
\begin{eqnarray}\label{eq:Si_ho}
S_1(x,y) &=& \frac{1}{6} (x^2+y^2+x\,y)\\
S_2(x,y) &=& \frac{1}{6} \\
S_3(x,y) &=& \frac{1}{180}(4x^2 + 4y^2 + 7 x\,y) \;,
\end{eqnarray}   
which agree with the results from a direct expansion of the exact propagator given in Eq. (\ref{eq:propag_hamonic_supp}) in $d=1$.

\vspace*{0.5cm}

{\it Bulk kernel and global density.} To compute the bulk kernel, we keep terms up to order ${\cal O}(t)$ in Eq. (\ref{eq:short_time}).  
We substitute this expansion in Eq. (\ref{eq:laplace2_supp}) and use the integral representation in Eq. (\ref{eq:bessel}). This gives 
\begin{eqnarray}\label{eq:kernel_gen}
K_\mu({\bf x},{\bf y})&=&\left({m\,[\mu -S_1({\bf x},{\bf y})]
\over 2 \pi^2 \hbar^2 ({\bf x}-{\bf y})^2}\right)^{d\over 4} J_{d\over 2}\left(\sqrt{2m({\bf x}-{\bf y})^2[\mu -S_1({\bf x},{\bf y})]\over \hbar^2}\right) \theta\left( 
\mu -S_1({\bf x},{\bf y})\right),
\end{eqnarray}
where $J_\nu(z)$ denotes a Bessel function of the first kind with index $\nu$. The support of the bulk part of the kernel is thus finite. The normalized particle density is given by $\rho_N({\bf x}) = K_\mu({\bf x},{\bf x})/N$, and the Fermi energy $\mu$ is determined from the normalization condition $\int d{\bf x} \, \rho_N({\bf x})=1
$ where the integral is over the support of the density. Using $S_1({\bf x},{\bf x})= V({\bf x})$ [see Eq. (\ref{S1})] gives the global density for arbitrary potential
\begin{equation}\label{eq:global}
\rho_N({\bf x}) \approx \frac{1}{N}\left(\frac{m}{2 \pi \hbar^2}\right)^{d/2} \frac{[\mu - V({\bf x})]^{d/2}}{\Gamma(d/2+1)} \;.
\end{equation}
At two generic points ${\bf x}$ and ${\bf y}$ in the bulk (far from the edges) with their separation $|{\bf x}- {\bf y}| \sim [N \rho_N({\bf x})]^{-1/d}$, Eq. (\ref{eq:kernel_gen}) simplifies to the scaling form 
\begin{eqnarray}\label{scaling_bulk_gen}
K_{\mu}({\bf x},{\bf y}) \approx \ell^{-d} {\cal K}_{\rm bulk}(|{\bf x}-{\bf y}|/\ell)
\end{eqnarray}
where $\ell = [N \rho_N({\bf x})\gamma_d]^{-1/d}$ is the typical separation in the bulk and {$\gamma_d =  \pi^{d/2} [\Gamma(d/2+1)]$}. The bulk scaling function is 
given explicitly by
\begin{eqnarray}\label{eq:kernel_bulk2_supp}
{\cal K}_{\rm bulk}(x) = \frac{J_{d/2}(2 x)}{(\pi x)^{d/2}}
\end{eqnarray}
as in Eq. (17) in the main text for the harmonic oscillator. The dependence on the potential $V({\bf x})$ enters only through the local density $\rho_N({\bf x})$ and hence through the scale factor $\ell$. However the scaling function associated with the bulk kernel in Eq. (\ref{eq:kernel_bulk2_supp}) is completely universal for all $V({\bf x})$.

\vspace*{0.5cm}

{\it Edge kernel.} We now focus on the behavior of the kernel near the edge of the global density where the edge is defined by $V({\bf r}_{\rm edge}) = \mu$. We consider $K_\mu({\bf r}_{\rm edge}+{\boldsymbol{a}}', {\bf r}_{\rm edge} +{\boldsymbol{b}}')$ where ${\boldsymbol{a}}'$ and ${\boldsymbol{b}}'$ are vectors such that $|{\boldsymbol{a}}'|,
 |{\boldsymbol{b}}'| \ll |{\bf r}_{\rm edge}| = r_{\rm edge}$ in a way we make more precise later. Henceforth, we focus only on spherically symmetric potential with a single minimum, e.g., {$V({\bf x}) = V(r)\sim r^p$, with $p>0$,
 in the region $r \sim r_{\rm edge}$ (no other scale in the potential)}. For such a potential, $r_{\rm edge} \sim \mu^{1/p}$. Substituting $V({\bf x}) = V(r)\sim r^p$ in the global density in Eq. (\ref{eq:global}) and using the normalization $\int d{\bf x} \rho_N({\bf x}) =1$, provides us with the estimate $\mu \sim N^{2p/(d(p+2))}$ for large $N$. Consequently, $r_{\rm edge} \sim N^{2/(d(p+2))}$ for large $N$. 
 
To proceed, we set ${\bf x} \equiv {\bf r}_{\rm edge} + {\bf a}'$ and ${\bf y} \equiv {\bf r}_{\rm edge} + {\bf b}'$ in Eqs. (\ref{S1}-\ref{S3}) and expand $S_1, S_2$ and $S_3$, assuming $|{\boldsymbol{a}}'|, |{\boldsymbol{b}}'| \ll {r}_{\rm edge}$. As in the case of the harmonic oscillator, only the terms of order ${\cal O}(t)$ and ${\cal O}(t^3)$ matter in the appropriate scaling regime. Keeping only these terms, the edge kernel is given by 
 \begin{equation}\label{eq:K_edge1}
K_\mu({\bf x}, {\bf y})\sim \left(m\over 2\pi \hbar\right)^{d\over 2} \int_\Gamma {dt\over 2\pi i} \frac{1}{t^{\frac{d}{2}+1}}  \exp\left[-{m\over 2\hbar t}\left(
{\boldsymbol{a}}'-{\boldsymbol{b}}'\right)^2\right]\exp\left[-{t\over 2\hbar}|\nabla V|(a'_n+b'_n) +{t^3\over 24m\hbar}|\nabla V|^2\right], 
 \end{equation}
where $\nabla V = \nabla V({\bf r}_{\rm edge})$ is evaluated at the edge. Here $a'_n$ and $b'_n$ denote the component of each vector normal to the edge. Following the analysis done for the harmonic oscillator, we then introduce the scaled dimensionless vectors ${\boldsymbol{a}}$ and ${\boldsymbol{b}}$ defined via ${\boldsymbol{a}}'= w_N\,{\boldsymbol{a}}$ and ${\boldsymbol{b}}'= w_N\,{\boldsymbol{b}}$, where the width $w_N$ has the dimension of length. Its precise value is fixed as follows. In order that both the terms of order ${\cal O}(t)$ and ${\cal O}(t^3)$ in Eq. (\ref{eq:K_edge1}) scale in the same way, 
 we choose $w_N$ as
 \begin{equation}\label{def_xi}
 w_N = {|\nabla V|^{-{1\over 3}}\hbar^{2\over 3}\over (2m)^{1\over 3}}\;.
 \end{equation}
Note that for the potentials {introduced above}
such that $V(r)\sim r^p$ (with $p>0$), then $|\nabla V| \sim r_{\rm edge}^{p-1}$ implying $w_N \sim N^{-\frac{2}{3d}(p-1)/(p+2)}$ for large $N$. Thus for $p>1$, the width shrinks with increasing $N$, while it increases for $0<p<1$. However, for any $p>0$, $w_N \ll r_{\rm edge}$ (recalling that $r_{\rm edge} \sim N^{2p/(d(p+2))}$). For the $d$-dimensional harmonic oscillator, $p=2$, Eq. (\ref{def_xi}) gives $w_N =  b_d\,N^{-1/(6d)}$ with $b_d =  [\Gamma(d+1)]^{-\frac{1}{6d}}/(\alpha \sqrt{2})$ -- in complete agreement with our previous analysis of the harmonic oscillator potential. {Going back to the integral (\ref{eq:K_edge1}) we find that it is controlled by 
a time scale, which we denote by $t_N$, such that terms $t^{-1}$, $t$ and $t^3$ are all of
order unity, which imply respectively that $t_N \sim \frac{m w_N^2}{\hbar} \sim \frac{\hbar}{w_N |\nabla V|} 
\sim \frac{(m \hbar)^{1/3}}{|\nabla V|^{2/3}}$ fully consistent with (\ref{def_xi}). We can thus
estimate the magnitude of the neglected terms. The term ${\cal O}(t^2)$, from Eq.~(\ref{S2}), 
scales as $\sim \frac{t^2}{m} \nabla^2 V$: it is thus
small iff $\frac{\hbar^2}{m w_N^2} \frac{\nabla^2 V}{|\nabla V|^2} \ll 1$, equivalently
$(\frac{\hbar^2}{m |\nabla V|^4})^{1/3} \nabla^2 V \ll 1$. If the potential does not
contain other scales, this term scales as $\sim r_{\rm edge}^{- \frac{p+2}{3}}$ and
is indeed negligible. Similarly it is easy to check that the neglected second term in $S_3$ in Eq.~(\ref{S3}) is indeed small since $\frac{\hbar^2 \nabla^4 V}{m |\nabla V|^2} \sim r_{\rm edge}^{-p-2}$.

Substituting the scaling variables in Eq. (\ref{eq:K_edge1}) and rescaling time appropriately (following the same procedure as in the harmonic oscillator case in section \ref{edge_kernel}) we get
 \begin{eqnarray}\label{eq:K_edge2_generalV}
 K_{\mu}({\bf x}, {\bf y})\approx \frac{1}{C_d \,w_N^d} \int_{\Gamma} \frac{d\tau}{2\pi i} \frac{1}{\tau^{\frac{d}{2}+1}} \, e^{-\frac{({\bf a} - {\bf b})^2}{2^{8/3}\tau} - \frac{(a_n + b_n)\tau}{2^{1/3}} + \frac{\tau^3}{3}} \;,
  \end{eqnarray}
where $C_d = (2^{\frac{4}{3}}\sqrt{\pi})^{d}$. For the $d$-dimensional harmonic oscillator, one can verify that Eq. (\ref{eq:K_edge2_generalV}) reduces exactly to the result in Eq. (\ref{eq:edge_kernel_supp}). Repeating the analysis done after Eq. (\ref{eq:edge_kernel_supp}), we finally get the scaling behavior of the edge kernel, 
\begin{eqnarray}\label{eq:def_edge_kernel}
K_{\mu}({\bf x},{\bf y}) \approx \frac{1}{w_N^d} {\cal K}_{\rm edge}\left(\frac{{\bf x} - {\bf r}_{\rm edge}}{w_N},\frac{{\bf y} - {\bf r}_{\rm edge}}{w_N}\right) \;,
\end{eqnarray}
where the scaling function is given explicitly by
\begin{eqnarray}\label{eq:K_edge3}
{\cal K}_{\rm edge}({\bf a},{\bf b}) = \int \frac{d^d q}{(2 \pi)^d}  e^{-i {\bf q} \cdot ({\bf a} - {\bf b})  } Ai_1\left(2^{\frac{2}{3}} q^2 + \frac{a_n+b_n}{2^{1/3}}\right)
\end{eqnarray}
where we recall that $Ai_1(z) = \int_z^\infty Ai(u) \, du$. Thus all the dependence on the potential $V(|{\bf x}|)$ is encoded in the width $w_N$ of the edge regime, but the scaling function ${\cal K}_{\rm edge}({\bf a},{\bf b})$ in Eq. (\ref{eq:K_edge3}) is universal, i.e., independent of $V(r)$. 

\vspace*{0.5cm}

{\it Edge density.} Finally, putting ${\bf a} = {\bf b}$ in Eq. (\ref{eq:K_edge3}) and using the definition $\rho_{\rm edge}({\bf x}) \equiv (1/N) K_\mu({\bf x},{\bf x}) \approx w_N^{-d} {\cal K}_{\rm edge}({\bf a}, {\bf a})/N$, we arrive at 
\beq\label{rho_edge_genV}
\rho_{\rm edge}({\bf x}) \approx \frac{1}{N} \frac{1}{w_N^d} \int \frac{d^d q}{(2\pi)^d} Ai_1(2^{2/3}\,(q^2+a)) \;,
\eeq
where $a = |{\bf a}|$ and $Ai_1(z) = \int_z^\infty Ai(x) \, dx$. To proceed, we rewrite Eq.~(\ref{rho_edge_genV}) as  
\beq\label{rho_edge_genV2}
\rho_{\rm edge}({\bf x}) \approx \frac{1}{N} \frac{\Gamma_d}{(2 \pi w_N)^d} \int_0^\infty dq\, q^{d-1} \int_{2^{2/3}(q^2+a)}^\infty Ai(z) \, dz \;,
\eeq
where $\Gamma_d = 2 \pi^{d/2}/\Gamma(d/2)$ is the surface area of the $d-$dimensional unit sphere. Making the change of variable $u=q^2 2^{2/3}$ and integrating by parts gives us
 \beq\label{rho_edge_genV3}
\rho_{\rm edge}({\bf x}) \approx \frac{1}{N} \frac{1}{w_N^d} F_d \left(\frac{r-r_{\rm edge}}{\, w_N} \right) \;,
\eeq  
where the scaling function $F_d(z)$ is given in Eq. (\ref{eq:identity3}) [which appears as Eq. (6) in the main text]. As in the case of the edge kernel, the dependence on $V(|{\bf x}|)$ appears only through the scale factor $w_N$, but the scaling function for the edge density $F_d(z)$ is universal.


\begin{thebibliography}{100}
\bibitem{footnote_potentials} Actually, the results hold for a wider class of potentials which are not necessarily spherically symmetric -- details will be published elsewhere \cite{us_long}.
\bibitem{foot_fermilevel} For finite $N$ note that the ground state can be degenerate with the Fermi level $\mu$ partially occupied. However, for large $N$, this degeneracy does not affect the asymptotic properties and one can assume that all states up to $\mu$ are fully populated.
\end{thebibliography}
\end{document}